\documentclass[11pt]{article}
\usepackage{amsmath}
\usepackage{amsfonts}
\usepackage{amssymb}
\usepackage{graphicx}
\usepackage{bm}
\usepackage{color}
\usepackage{amscd}

\def\bea{\begin{eqnarray}}
\def\eea{\end{eqnarray}}

\begin{document}
\begin{center}
\LARGE {\bf On the Hamiltonian Analysis of Spin-3 Chern-Simons-Like Theories of Gravity }
\end{center}

\begin{center}
{M. R. Setare \footnote{E-mail: rezakord@ipm.ir}\hspace{1mm} ,
H. Adami \footnote{E-mail: hamed.adami@yahoo.com}\hspace{1.5mm} \\
{\small {\em  Department of Science, University of Kurdistan, Sanandaj, Iran.}}}\\

\end{center}

\begin{center}
{\bf{Abstract}}\\In this paper, we consider spin-3 Chern-Simons-like theories of gravity as extended theories of spin-3 gravity in (2+1)- dimension. 
In order to determine the number of local degrees of freedom we present the Hamiltonian formulation of these theories.
We extract the Hamiltonian density, then we find primary and secondary constraints of these theories. Then we obtain the Poisson brackets of the primary and the secondary constraints. After that we count the number of local degrees of freedom of spin-3 Chern-Simons-like theories of gravity. We apply this method on spin-3 Einstein-Cartan gravity and spin-3 topologically massive gravity. According to the our results the spin-3 Einstein-Cartan gravity and the spin-3 topologically massive gravity have respectively zero and one bulk local degree of freedom. 
\end{center}

\section{Introduction}
It is well known that pure Einstein-–Hilbert gravity in three dimensions exhibits no propagating physical degrees of freedom \cite{2',3'}. But adding the gravitational Chern-Simons term produces a
propagating massive graviton \cite{4'}. The resulting theory
is called topologically massive gravity (TMG).  A new theory of massive gravity (NMG) in three dimensions
has been proposed by Bergshoeff at. al some years ago \cite{2}.  Both TMG and NMG have a bulk-boundary
unitarity conflict, In another term either the bulk or the boundary theory is non-unitary, so there is a clash between the positivity of the two Brown-Henneaux boundary $c$ charges and the bulk energies. Recently an interesting three dimensional massive gravity introduced by Bergshoeff, et. al \cite{3} which dubbed Minimal Massive Gravity (MMG), which has the
same minimal local structure as TMG. The MMG model has the same gravitational
degree of freedom as the TMG has and the linearization of the metric field equations for MMG yield a single propagating
massive spin-2 field. It seems that the single massive degree of freedom of MMG is
unitary in the bulk and gives rise to a unitary CFT on the boundary. Following this paper some interesting works have been done on MMG model \cite{18}. More recently the author of \cite{4} has introduced  Generalized Minimal Massive Gravity (GMMG), an interesting modification of MMG. GMMG is a unification of MMG with New Massive Gravity (NMG), so this model is realized by adding higher-derivative deformation term to the Lagrangian of MMG. As has been shown in \cite{4}, GMMG also avoids the aforementioned ``bulk-boundary unitarity clash''.
Calculation of the GMMG action to quadratic order about $AdS_3$ space show that the theory is free of negative-energy bulk modes. Also Hamiltonian analysis show that the GMMG model has no Boulware-Deser ghosts and this model propagate only two physical modes.
 So these models are viable candidate for semi-classical limit of a unitary quantum $3D$ massive gravity.\\
 Generally these types of theories in $2+1-$dimension (e.g. Topological massive gravity (TMG), New massive gravity (NMG), Minimal massive gravity (MMG), Generalized minimal massive gravity (GMMG), etc), called the Chern-Simons-like theories of gravity \cite{1}. In another term one can write the Lagrangian of these types of theories as a Lagrangian 3-form constructed from one-form
fields and their exterior derivatives.\\
Higher spin gravity in its simplest
form is an extension of ordinary gravity that includes a
massless scalar and massless fields with spins $S=3,4,...$ \cite{0}. In \cite{00} Vasiliev proposed a system of gauge invariant nonlinear dynamical equations for totally symmetric massless fields of all spins in (A)dS backgrounds. Higher-spin theories in $AdS_3$, like ordinary gravity, they possess no propagating degrees of freedom \cite{5,3''}. The authors of \cite{12} have done the generalization of topologically massive gravity to
higher spins, specifically spin-3 (see also \cite{7'}). In recent paper \cite{30} we generalized the spin-3 gravity in the first order formalism to the Spin-3 Chern-Simons-like theories of gravity (Spin-3 CSLTG). It would be interesting to study a Hamiltonian analysis of Spin-3 CSLTG as the authors of \cite{1} have done for ordinary CSLTG. This is our aim in this paper. It is important that by a Hamiltonian analysis one can obtain the number of local degrees of freedom exactly and independent of a linearized approximation.  For example a Hamiltonian analysis  show that the GMMG model has no Boulware-Deser ghosts and this model propagate only two physical modes \cite{4}. So, in this paper we present the Hamiltonian formulation of Spin-3 Chern-Simons-like theories of gravity. Then by this Hamiltonian approach we could determine the number of local degrees of freedom.
By applying our method to spin-3 Einstein-Cartan gravity and spin-3 topologically massive gravity we show that similar to the ordinary cases of these theories, spin-3 Einstein-Cartan gravity do not propagate any local bulk degree of freedom, but spin-3 topologically massive gravity propagate only one bulk local degree of freedom. 
\section{Spin-3 Chern-Simons-Like Theories of Gravity}
The ordinary chern-Simons-like theories of gravity are investigated in some papers, for instance see \cite{4',2,3,4,1}. These theories can be generalize to spin-3 one (Spin-3 CSLTG) by the following Lagrangian \cite{30}
\begin{equation}\label{1}
  L = tr \{ \frac{1}{4} \tilde{g}_{rs} a^{r} \wedge d a^{s} + \frac{1}{6} \tilde{f}_{rst} a^{r} \wedge a^{s} \wedge a^{t} \},
\end{equation}
where $ a^{r} = a^{rA}_{\hspace{3 mm} \mu} J_{A} dx^{\mu} $ are $sl(3,\mathbb{R})$ Lie algebra valued one-forms, $J _{A}$ denotes the generators of $sl(3,\mathbb{R})$ algebra. The algebra $sl(3,\mathbb{R})$ have eight generators, so $A=1, \dots , 8$.  In above equation, $r=1,...,N$ refers to flavour index, $\tilde{g} _{rs}$ is a symmetric constant metric on the flavour space and $\tilde{f} _{rst}$ is a totally symmetric "flavour tensor" which is interpreted as coupling constant. We take $ a^{r} =\{ e, \omega , h , \cdots \} $, where $e$, $\omega$ and $ h $ are generalized dreibein, generalized spin-connection and auxiliary field, respectively.\\
The Killing form in the fundamental representation of $sl(3, \mathbb{R})$ is defined as
\begin{equation}\label{2}
  K_{AB}=\frac{1}{2} tr(J_{A} J_{B}).
\end{equation}
Anti-symmetric structure constants of the Lie algebra is given by
\begin{equation}\label{3}
  f_{ABC}=\frac{1}{2} tr([ J_{A} , J_{B} ]J_{C}) ,
\end{equation}
so, we can rewrite the Spin-3 CSLTG Lagrangian as
\begin{equation}\label{4}
  L =  \frac{1}{2} \tilde{g}_{rs} K_{AB} a^{rA} \wedge d a^{sB} + \frac{1}{6} \tilde{f}_{rst} f_{ABC} a^{rA} \wedge a^{sB} \wedge a^{tC}.
\end{equation}
It is easy to check that this Lagrangian reduces to the ordinary CSLTG one when we suppress the spin-3 field contribution. For ordinary CSLTG case, we have $K_{AB} \equiv \eta _{a b}$ and $f_{ABC} \equiv \varepsilon _{a b c}$, where $a,b,c, \cdots$ denote Lorentz indices and $\eta _{ab}$ is simply the Minkowski metric.
\section{Hamiltonian Analysis}
One can separate space and time  part of $a^{rA}$ in the following form
\begin{equation}\label{5}
  a^{rA} = a^{rA}_{\hspace{3.5 mm} 0 } dt + a^{rA}_{\hspace{3.5 mm} i } dx^{i},
\end{equation}
which leads to the Lagrangian density
\begin{equation}\label{6}
  \mathcal{L} = - \frac{1}{2} \varepsilon ^{ij} \tilde{g}_{rs} K_{AB} a^{rA}_{\hspace{3.5 mm} i } \dot{a}^{sB} _{\hspace{3.5 mm} j } - a^{rA}_{\hspace{3.5 mm} 0 } \phi _{rA}.
\end{equation}
In the above equation the dot on the $a^{sB} _{\hspace{3.5 mm} j }$ denotes the time derivative, and $\varepsilon ^{ij} = \varepsilon ^{0ij}$, also $\phi _{rA}$ defines as
\begin{equation}\label{7}
  \phi _{rA} = - \varepsilon ^{ij} \tilde{g}_{rs} K_{AB} \partial _{i} a^{sB} _{\hspace{3.5 mm} j } - \frac{1}{2} \varepsilon ^{ij} \tilde{f}_{rst} f_{ABC} a^{sB} _{\hspace{3.5 mm} i } a^{tC} _{\hspace{3.5 mm} j }.
\end{equation}
We can read off the Hamiltonian density from the Lagrangian density \eqref{6} as
\begin{equation}\label{8}
  \mathcal{H} = a^{rA}_{\hspace{3.5 mm} 0 } \phi _{rA},
\end{equation}
this is the Hamiltonian of a constraint system. Therefore, we follow Dirac's procedure \cite{7} to find the primary and secondary constraints and then we obtain their Poisson brackets.
\section{Primary and Secondary Constraints}
Since the Lagrangian \eqref{6} does not depend on time derivative of $a^{rA}_{\hspace{3.5 mm} 0 }$, we can interpret $\phi _{rA}=0$ as the primary constraints and $a^{rA}_{\hspace{3.5 mm} 0 }$ as Lagrange multipliers. Thus, we have $8N$ primary constraints. The Poisson brackets of the canonical variables are
\begin{equation}\label{9}
  \{ a^{rA}_{\hspace{3.5 mm} i} (x) , a^{sB}_{\hspace{3.5 mm} j} (y) \}_{P.B} = \varepsilon _{ij} \tilde{g} ^{rs} K ^{AB} \delta ^{(2)} (x-y),
\end{equation}
which can be determined by inverting the first term of the Lagrangian \eqref{6}. To calculate the Poisson brackets of the primary constraint
functions, we define the "smeared" functions associated to the primary constraints
\begin{equation}\label{10}
  \Phi [\xi] = \int_{\Sigma} d ^{2} x \hspace{1.5 mm} \xi ^{rA}(x) \phi_{rA}(x),
\end{equation}
where $\Sigma$ is a space-like hypersurface and $\xi ^{rA}(x)$ is a test function. Varying the above smeared function  with respect to the fields $a^{rA} _{\hspace{3.5 mm} i} $ gives
\begin{equation}\label{11}
\begin{split}
   \delta \Phi [\xi] = & \int_{\Sigma} d ^{2} x \varepsilon ^{ij} \left\{ \tilde{g} _{rs} K _{AB} \partial _{i} \xi ^{rA} + \tilde{f}_{rst} f_{ABC} a^{tC} _{\hspace{3.5 mm} i } \right\} \delta a^{sB} _{\hspace{3.5 mm} j } \\
     & - \int_{\partial \Sigma} d x ^{i} \tilde{g} _{rs} K _{AB} \xi ^{rA} \delta a^{sB} _{\hspace{3.5 mm} i }.
\end{split}
\end{equation}
It is obvious that the functionals $\Phi [\xi]$ are not differentiable, but one can add a term to the smeared function to define a new function $\varphi [\xi] = \Phi [\xi] + Q [\xi]$ such that their variation with respect to $a^{rA} _{\hspace{3.5 mm} i} $ do not provide boundary terms, where $Q[\xi]$ is given by
\begin{equation}\label{12}
  Q[\xi] = \int_{\partial \Sigma} d x ^{i} \tilde{g} _{rs} K _{AB} \xi ^{rA} a^{sB} _{\hspace{3.5 mm} i },
\end{equation}
hence, we have
\begin{equation}\label{13}
  \varphi [\xi] = \int_{\Sigma} d^{2}x \hspace{1.5 mm} \varepsilon ^{ij} \left\{ \tilde{g}_{rs} K_{AB} \partial _{i} \xi^{rA} a^{sB} _{\hspace{3.5 mm} j } - \frac{1}{2} \tilde{f}_{rst} f_{ABC} \xi^{rA} a^{sB} _{\hspace{3.5 mm} i } a^{tC} _{\hspace{3.5 mm} j } \right\}.
\end{equation}
Now, we can calculate the Poisson brackets of the constraint functions
\begin{equation}\label{14}
\begin{split}
\{ \varphi [\alpha],\varphi [\beta] \}_{P.B.}= & \varphi [[\alpha,\beta]] - \int_{\Sigma} d^{2}x \alpha^{r}_{\hspace{1.5 mm} A} \beta ^{s}_{\hspace{1.5 mm} B} (\mathcal{P}^{AB})_{rs} \\
& - \int_{\partial \Sigma} dx^{i} g_{rs} K_{AB} \alpha^{rA} \partial _{i} \beta ^{sB},
\end{split}
\end{equation}
where
\begin{equation}\label{15}
(\mathcal{P}^{AB})_{rs}=f^{t}_{\hspace{1.5 mm} q[r } f_{s]pt} K ^{AB} \Delta ^{pq} + 2 f^{t}_{\hspace{1.5 mm} r[s} f_{q]pt} (V^{AB})^{pq} ,
\end{equation}
\begin{equation}\label{16}
\Delta ^{pq}=\varepsilon ^{ij} K_{AB} a^{pA}_{\hspace{1.5 mm} i} a^{qB}_{\hspace{1.5 mm} j}, \hspace{1.5 cm} (V^{AB})^{pq} = \varepsilon ^{ij} a^{pA}_{\hspace{2.5 mm} i} a^{qB}_{\hspace{2.5 mm} j} ,
\end{equation}
and
\begin{equation}\label{17}
  [\alpha , \beta]^{t}_{\hspace{1.5 mm} C} = \tilde{f}^{t} _{\hspace{1.5 mm} rs} f_{ABC} \alpha ^{rA} \beta ^{sB}.
\end{equation}
One can choose $\xi ^{rA}$ in a way that the boundary integrals vanish. Then using Eq.\eqref{14} and definition of $\varphi [\xi]$, we find the Poisson brackets of the primary constraints
\begin{equation}\label{18}
\{ \phi _{r}^{\hspace{1.5 mm} A }(x),\phi _{s}^{\hspace{1.5 mm} B }(y) \} _{P.B.}=\delta ^{(2)}(x-y) [ \tilde{f}^{t}_{\hspace{1.5 mm} rs } f ^{AB} _{\hspace{4.5 mm} C} \phi _{t}^{\hspace{1.5 mm} C } - (\mathcal{P}^{AB})_{rs}] .
\end{equation}
The consistency conditions which guarantee time-independence of the primary constraints are
\begin{equation}\label{19}
  \frac{d}{dt} \phi _{s} ^{\hspace{1.5 mm} A}(x) = \{ \mathcal{H} (y) , \phi _{s} ^{\hspace{1.5 mm} A} (x) \} \approx - a^{r}_{\hspace{1.5 mm} B 0} (\mathcal{P}^{BA})_{rs} \delta ^{(2)}(x-y) \approx 0.
\end{equation}
Equations of motion arise from the Lagrangian \eqref{4} are
\begin{equation}\label{20}
   \tilde{g}_{rs} K_{AB} d a^{sB} +\frac{1}{2} \tilde{f}_{rst} f_{ABC} a^{sB} \wedge a^{tC} =0,
\end{equation}
Taking the exterior derivative of this equation, we have following integrability conditions 
\begin{equation}\label{21}
  \tilde{f} ^{t}_{\hspace{1.5 mm} q [r} \tilde{f}_{s]pt} K_{BC} a^{rA} \wedge a^{pB} \wedge a^{qC}=0.
\end{equation}
We use the space-time decomposition \eqref{5}, then we find that
\begin{equation}\label{22}
  \tilde{f} ^{t}_{\hspace{1.5 mm} q [r} \tilde{f}_{s]pt} K_{BC} a^{rA} \wedge a^{pB} \wedge a^{qC}=a^{r}_{\hspace{1.5 mm} B 0} (\mathcal{P}^{BA})_{rs} =0.
\end{equation}
Thus, consistency conditions are equivalent to a set of integrability conditions. Now, we consider Eq.\eqref{22} in the following form
\begin{equation}\label{23}
  \tilde{f} ^{t}_{\hspace{1.5 mm} q [r} \tilde{f}_{s]pt} K_{BC} a^{rA}_{\hspace{3.5 mm} \lambda} a^{pB}_{\hspace{3.5 mm} \mu} a^{qC}_{\hspace{3.5 mm} \nu} dx^{\lambda} \wedge dx^{\mu} \wedge dx^{\nu}=0
\end{equation}
If the sum over $r$ is non-zero for only one value of $r$, so in order that the solution of the above equations becomes independent from $a^{rA} _{\hspace{3.5 mm} 0} $ for a fixed $r$, we must have
\begin{equation}\label{24}
  \tilde{f} ^{t}_{\hspace{1.5 mm} q [r} \tilde{f}_{s]pt} K_{BC} a^{pB}_{\hspace{3.5 mm} i} a^{qC}_{\hspace{3.5 mm} j} dt \wedge dx^{i} \wedge dx^{j}=0,
\end{equation}
or equivalently
\begin{equation}\label{25}
  \tilde{f} ^{t}_{\hspace{1.5 mm} q [r} \tilde{f}_{s]pt} \Delta ^{pq}=0.
\end{equation}
In this way, arbitrariness of $a^{rA} _{\hspace{3.5 mm} 0}$ to be preserved and therefore the secondary constraints can be expressed as
\begin{equation}\label{26}
  \psi _{I} \equiv F_{I,pq} \Delta ^{pq} = \tilde{f} ^{t}_{\hspace{1.5 mm} q [r} \tilde{f}_{s]pt} \Delta ^{pq}, \hspace{0.7 cm} I=1,\cdots,M,
\end{equation}
where $M$ is the number of secondary constraints. Now, we calculate the Poisson brackets of the secondary and the primary constraints and we obtain
\begin{equation}\label{27}
\{ \varphi[\xi],\psi _{I} \}_{P.B.}=2 \varepsilon ^{ij} \left( F_{I,rp} K_{AB} \partial _{i} \xi ^{rA} a^{pB}_{\hspace{3.5 mm} j } + \tilde{f}^{t}_{\hspace{1.5 mm} rs } F_{I,pt} f_{ABC} \xi ^{rA} a^{sB}_{\hspace{3.5 mm} i } a^{pC}_{\hspace{3.5 mm} j } \right),
\end{equation}
then we find that
\begin{equation}\label{28}
\begin{split}
\{ \phi _{r}^{\hspace{1.5 mm} A }(x) ,\psi _{I}(y) \}_{P.B.}= & 2 \varepsilon ^{ij} (-F_{I,rp} \partial _{i} a^{pA}_{\hspace{3.5 mm} j } \\
& +\tilde{f}^{t}_{\hspace{1.5 mm} rs } F_{I,pt} f^{A}_{\hspace{2.5 mm} BC} a^{sB}_{\hspace{3.5 mm} i } a^{pC}_{\hspace{3.5 mm} j } ) \delta ^{(2)}(x-y).
\end{split}
\end{equation}
In a similar way, we find the Poisson brackets of the secondary constraints as
\begin{equation}\label{29}
\{ \psi _{I}(x),\psi _{J}(y) \}_{P.B}=-4 F_{I,pq} F_{J,rs} \tilde{g}^{qs} \Delta ^{pr} \delta ^{(2)}(x-y).
\end{equation}
By similar arguments for ordinary CSLTG \cite{1}, we can find the number of local degrees of freedom by the following formula
\begin{equation}\label{30}
D= 16 N - 2 \times (8 N - rank ( \mathcal{P} ) - M )-1 \times ( rank( \mathcal{P} ) + 2 M )=rank ( \mathcal{P} ).
\end{equation}
It should be noted that this equation is valid when the Poisson brackets of the secondary constraints all vanish.
\section{Examples}
Now, we use this procedure to determine the number of local degrees of freedom of following models.
\subsection{Spin-3 Einstein-Cartan Gravity}
 A three dimensional spin-3 gravity theory is given by following action,
 \begin{equation}\label{310}
  S_{EH}= \frac{1}{16 \pi G} \int ( e \wedge \mathcal{R} +\frac{1}{3l^{2}} e \wedge e \wedge e ),
\end{equation}
where $\mathcal{R}$ is the generalized curvature 2-form which is as following
\begin{equation}\label{160}
  \mathcal{R} = d \omega + \omega \wedge \omega \hspace{0.5 cm} \Leftrightarrow
  \hspace{0.5 cm} \mathcal{R} ^{A} = d \omega ^{A} + \frac{1}{2} f ^{A} \hspace{0.1 mm} _{BC} \omega ^{B} \wedge \omega ^{C} .
\end{equation}
For the above spin-3 Einstein-Cartan gravity, the non-zero components of the flavour metric and the flavour tensor are
\begin{equation}\label{31}
  \tilde{g}_{e \omega}=-1, \hspace{0.7 cm} \tilde{f}_{eee}=\Lambda , \hspace{0.7 cm} \tilde{f}_{e \omega \omega}=-1.
\end{equation}
In this model, we have
\begin{equation}\label{32}
  \tilde{f}^{t}_{\hspace{1.5 mm} q[r } \tilde{f}_{s]pt}=0,
\end{equation}
and therefore all components of $(\mathcal{P}^{AB})_{rs}$ are zero. Thus, as one can expect, Einstein-Cartan gravity in three dimension has no local degrees of freedom. In 3-dimension the Fronsdal
gauge fields with spin $S>1$ do not propagate local degree of freedom \cite{5}. The Poisson brackets of the primary constraints for this model reduce to
\begin{equation}\label{33}
\{ \phi _{r}^{\hspace{1.5 mm} A }(x),\phi _{s}^{\hspace{1.5 mm} B }(y) \} _{P.B.}= \tilde{f}^{t}_{\hspace{1.5 mm} rs } f ^{AB} _{\hspace{4.5 mm} C} \phi _{t}^{\hspace{1.5 mm} C } \delta ^{(2)}(x-y).
\end{equation}
In this model, all of the primary constraints are first class constraints.
\subsection{Spin-3 Topologically Massive Gravity}
Now we consider a Lagrangian describing the spin-3 field coupled to TMG \cite{12,7'}. The Lagrangian of the spin-3 Topologically Massive Gravity is given by
\begin{equation}\label{340}
  \mathcal{L}= tr \{ - \sigma e \wedge \mathcal{R} + \frac{\Lambda}{3} e \wedge e\wedge e + \frac{1}{2 \mu} \left(\omega d \omega + \frac{2}{3} \omega \wedge \omega \wedge \omega \right) + h \wedge \mathcal{T} \}.
\end{equation}
In the above Lagrangian $h$ is an auxiliary $sl(3,\mathbb{R})$ Lie algebra valued one-form field. Also $\mu$ is a mass parameter, $\sigma$ is a sign and $\Lambda$ denotes the cosmological parameter. $\mathcal{T}$ is the generalized torsion 2-form as
\begin{equation}\label{15}
  \mathcal{T}= e_{\lambda} \Gamma ^{\lambda} _{\mu \nu} dx^{\mu} \wedge dx^{\nu}.
\end{equation}
where $ \Gamma ^{\lambda} _{\mu \nu} $ can be interprets as Affine connection.
For the spin-3 topologically massive gravity, the non-zero components of the flavour metric and the flavour tensor are
\begin{equation}\label{34}
  \begin{split}
       & \tilde{g}_{e \omega}= - \sigma , \hspace{0.7 cm} \tilde{g}_{e h}= 1 , \hspace{0.7 cm}\tilde{g}_{\omega \omega}= \frac{1}{\mu},   \\
       & \tilde{f}_{e \omega \omega}= - \sigma , \hspace{0.7 cm} \tilde{f}_{e h \omega}= 1 , \hspace{0.7 cm}\tilde{f}_{\omega \omega \omega}= \frac{1}{\mu} , \hspace{0.7 cm} \tilde{f}_{eee}=\Lambda.
  \end{split}
\end{equation}
One can show that the non-zero components of $f^{t}_{\hspace{1.5 mm} q[r } f_{s]pt}$ in this model are given by
\begin{equation}\label{35}
   \tilde{f}^{t}_{\hspace{1.5 mm} h[e } \tilde{f}_{h]et} = \tilde{f}^{t}_{\hspace{1.5 mm} e[h } \tilde{f}_{e]ht}= \frac{\mu}{2}.
\end{equation}
In this model, integrability conditions \eqref{21} become
\begin{equation}\label{36}
  K_{BC} e^{A} \wedge e^{B} \wedge h^{C}=0
\end{equation}
and hence we have one secondary condition
\begin{equation}\label{37}
  \psi_{1} = \Delta ^{eh} \approx 0.
\end{equation}
By substituting Eq.\eqref{35} and Eq.\eqref{37} into Eq.\eqref{15}, we have
\begin{equation}\label{38}
  (\mathcal{P}^{AB})_{rs} = - \mu \left(
                              \begin{array}{ccc}
                                (V^{AB})^{hh} & 0 & (V^{BA})^{eh}\\
                                0 & 0 & 0 \\
                                -(V^{AB})^{eh} & 0 & (V^{AB})^{ee} \\
                              \end{array}
                            \right)
\end{equation}
By some calculations with maple, we find that the rank of $\mathcal{P}$ is two. Therefore Spin-3 TMG model in non-linear regime has one bulk local degrees of freedom.

\section{Conclusion}
We have considered Spin-3 Chern-Simons-like theories of gravity. The Lagrangian of such theories is given by Eq.\eqref{1}. As has been discussed in \cite{1} the Chern-Simons formulation of gravity models is well-adapted to a Hamiltonian analysis. Hamiltonian formulation allows to count the number of local degrees of freedom of these theories in the non-linear regime. We obtained Hamiltonian density corresponds to these theories  by Eq.\eqref{8}. We deduced that Spin-3 CSLTG have $8N$ primary constraints, where $N$ is number of the flavour indices. We calculated the Poisson brackets of the primary constraints which are given by Eq.\eqref{18}. We have shown that the consistency conditions \eqref{19} are equivalent to the integrability conditions \eqref{21}. Using this result, we found the secondary constraints \eqref{26}. Also, we calculated the Poisson brackets of the primary and secondary constraints, see Eq.\eqref{28} and Eq.\eqref{29}. By similar arguments for ordinary CSLTG \cite{1}, we found the number of local degrees of freedom by Eq.\eqref{30}. As examples, we applied this method to the spin-3 Einstein-Cartan gravity and the spin-3 topologically massive gravity in order to count the number of local degrees of freedom of  these theories. According to the our results  the number of  bulk local degrees of freedom of these models are zero and one, respectively. These results are matched with results of papers \cite{5,3'',12,7'}. Simply one can apply our method to the other interesting Spin-3 Chern-Simons-like theories of gravity, such as Spin-3 MMG, and Spin-3 GMMG.

\end{document}